# Observation of Klein bottle quadrupole topological insulators in electric circuits

Xizhou Shen, [1, §] Keyu Pan, [1, §] Xiumei Wang, [2*] and Xingping Zhou[3*]

[1] *College of Integrated Circuit Science and Engineering, Nanjing University of Posts and Telecommunications, Nanjing 210003, China*

[2] *College of Electronic and Optical Engineering, Nanjing University of Posts and Telecommunications, Nanjing 210003, China*

[3] *Institute of Quantum Information and Technology, Nanjing University of Posts and Telecommunications, Nanjing 210003, China*

*wxm@njupt.edu.cn

*zxp@njupt.edu.cn

*§ These authors contributed equally to this work.*

The Klein bottle Benalcazar-Bernevig-Hughes (BBH) insulator phase plays a pivotal role in understanding higher-order topological phases. The insulator phase is characterized by a unique feature: a nonsymmorphic glide symmetry that exists within momentum space, rather than real space. This characteristic transforms the Brillouin zone's fundamental domain into a structure of Klein bottle. Here, we report an observation of a Klein bottle topoelectrical model under $\mathbb{Z}_2$ gauge fields. To provide a comprehensive understanding of the different corner distributions of odd and even unit cells, we present theoretical calculations and demonstrate that the symmetry properties significantly affect the topological nature. These theoretical predictions are confirmed by experimental results, which demonstrate the practical feasibility of such topological configurations in electronic circuits. Our work establishes a vital connection between the realms of condensed matter physics and circuit systems, thereby paving a pathway for investigating exotic condensed matter physics.

The field of topological matter originates from the discovery of the quantum hall effect [1-6] and the Chern insulator [7-11], which serve as fundamental pillars in our current comprehension of topological phases. The quantum hall effect manifests in two-dimensional (2D) electron systems under strong magnetic fields discovered by Klaus von Klitzing [4], thereby unveiling the phenomenon of quantized conductance. Following this discovery, F. D. M. Haldane discovers the Chern insulator [8]. It is associated with the quantum hall effect but does not require an external field and exhibits topologically protected edge states. These developments have historically been framed within the study of topological phases characterized by the orientable Brillouin torus.

The advent of gauge symmetry [12-14] marks a significant transformation in the realm of topological matter. It enhances the diversity of topological phases, thereby expediting the development of higher-order topological insulators [15-19]. The Benalcazar-Bernevig-Hughes (BBH) model [20-22] has played a pivotal role in advancing the study of topological phases. These topological insulators are identified by their quantized quadrupole moment, which is enforced by symmetries, such as chiral symmetry. The evolution of the 2D momentum-space Brillouin zone (BZ) shifts from a traditional orientable double-torus to a non-orientable Brillouin Klein bottle [21, 23-34]. This shift has revealed new topological constructs, particularly Klein bottle BBH insulators [21], which are characterized by their unique corner states. Despite the theoretical establishment of Klein bottle BBH insulators, their experimental realization has remained a challenge, with only phononic crystal systems providing tangible evidence of these properties [26-28, 31]. Although many topological phases have been implemented by circuits before [35-38], a circuitry-based Klein bottle has not been explored.

In this work, we construct a Klein bottle BBH topoelectrical model entirely through passive components (only using capacitors and inductors). Leveraging this model, our work reveals the key attribute of nonsymmorphic glide symmetry in momentum space. The discovery transforms the topological configuration of the model into a Klein bottle. We seek their zero-energy states by conducting a comprehensive analysis of the band structures for both odd and even unit cells. The unique properties of corner states in odd and even unit cells are evident, highlighting the presence of higher-order topological insulators. This observation underscores the non-trivial topological of our system. We also propose an odd-even switching mechanism on a single printed circuit board (PCB). This methodology allows for the direct observation and analysis of corner state characteristics across different unit cell configurations. Additionally, our mathematical analysis demonstrates that the symmetry properties of odd and even unit cells significantly affect the system's topological characteristics.

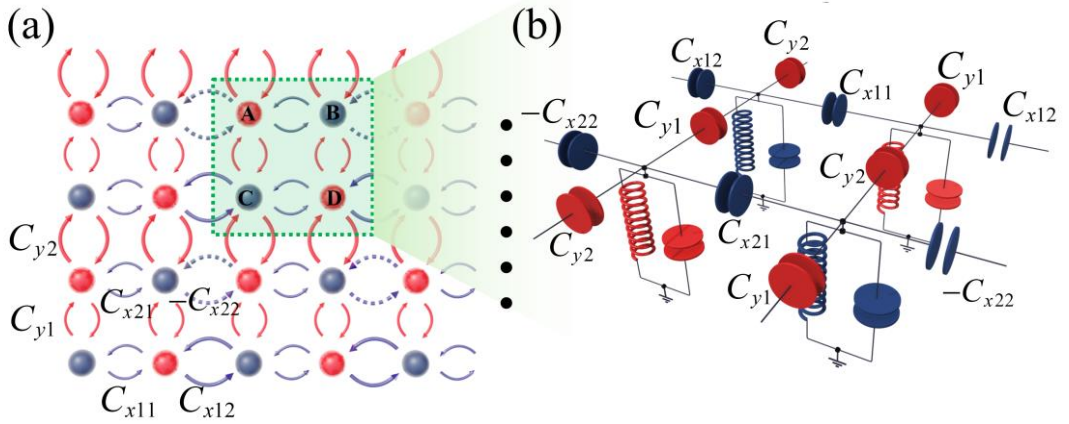

FIG. 1. Klein bottle BBH topoelectrical structure. (a) The extended lattice model of the Klein bottle BBH topoelectrical structure. Dark blue arrows indicate the hopping amplitude in the *x*-direction, while red arrows represent the *y*-direction, and dashed lines denote a negative sign. (b) The specific circuit model.

We start by considering the translation symmetry $L_x$ along the *x*-direction, and the reflection symmetry $M_y$ that inverses the *y*-direction. Usually, the spin free systems always have $[L_x, M_y] = 0$ and $L_x M_y L_x^{-1} M_y^{-1} = 1$ [21, 23, 31]. Recent groundbreaking theoretical advancements have demonstrated that artificially imposed gauge fields can fundamentally transform the conventional two-dimensional (2D) BZ, typically represented as an orientable double-torus [23]. This transformation results in a non-orientable Brillouin Klein bottle, which possesses entirely distinct topological properties and new topological phases. In this scenario, we impose the imposition of a $\mathbb{Z}_2$ gauge field. For any given gauge configuration, an equivalent configuration's symmetry operator $\mathsf{R} = \mathsf{G}R$ is defined by the combination of the original configuration's operator $R$ and a gauge transformation $\mathsf{G}$. It is necessary, although the flux configuration remains invariant under R, the specific gauge connection configuration may undergo alterations due to R. The restoration of the original gauge connections necessitates an additional gauge transformation, which is determined by assigning a +1 or -1 sign to each position in the basis. Here, we have $\mathsf{M}_x = \mathsf{G}M_x$, $\mathsf{L}_y = \mathsf{G}L_y$ [31], which make

$$\begin{aligned}
-1 &= \mathsf{M}_x \mathsf{L}_y \mathsf{M}_x^{-1} \mathsf{L}_y^{-1} = \mathsf{G}_M M_x \mathsf{G}_y L_y (\mathsf{G}_M M_x)^{-1} (\mathsf{G}_y L_y)^{-1} \\
&= \mathsf{G}_M (M_x \mathsf{G}_y M_x^{-1}) M_x L_y M_x^{-1} L_y^{-1} (L_y \mathsf{G}_M L_y^{-1}) \mathsf{G}_y^{-1} \\
&= \mathsf{G}_M(i) \mathsf{G}_y(M_x(i)) \mathsf{G}_M^*(L_y^{-1}(i)) \mathsf{G}_y^*(i)
\end{aligned} \quad (1)$$

To represent the relationships between the hopping amplitudes for two quadrupole units (unit cells) in *x*-direction, we can start by noting that the order of the four nodes in a clockwise direction from the top right corner is 1-4 and get $t_{23} = t_{14} \mathsf{G}_M(2) \mathsf{G}_M^*(3)$, $t_{43} = t_{12} \mathsf{G}_M(4) \mathsf{G}_M^*(3)$. Then the calculated flux is

$$\begin{aligned}
e^{-i\Phi} &= e^{i\phi_{12}} e^{i\phi_{23}} e^{i\phi_{34}} e^{i\phi_{41}} = e^{i\phi_{12}} [e^{i\phi_{14}} \mathsf{G}_M(2) \mathsf{G}_M^*(3)][e^{i\phi_{21}} \mathsf{G}_y(3) \mathsf{G}_y^*(4)] e^{i\phi_{41}} \\
&= \mathsf{G}_M^*(3) \mathsf{G}_y^*(M_x(3)) \mathsf{G}_M(L_y^{-1}(3)) \mathsf{G}_y(3) = (-1)^* = -1
\end{aligned} \quad (2)$$

where $\Phi$ is the flux. The result "-1" in Eq. (2) represents the flux as π. When a $\mathbb{Z}_2$ gauge flux of π passes through a quadrupole unit, it transports a particle from position 1 to position 3 along both the *x* and *y* axes. However, each path incorporates a π flux. Consequently, this configuration leads to an anticommutation relationship between the translation symmetry and the reflection symmetry, described as $\{M_y, L_x\} = 0$.

We initially present a 2D tight-binding model as depicted in Fig. 1(a), where each quadrupole contains four sites. The hopping phases are set to 1 or -1, represented by solid or dashed lines, respectively. This configuration results in each plaquette having a flux of either 0 or π. For a unit indicated by green box, we construct the topology through an electrical circuit as shown in Fig. 1(b). Each node corresponds to an LC resonator with one end connected to the ground, while the other coupled to adjacent resonators via capacitors. Positive hops represented by solid lines are achieved using positive capacitors, whereas negative hops represented by dashed lines are achieved using negative capacitors (achieved by inductances). In this model, $\tau$ and $\sigma$ represent Pauli matrices, corresponding to different positional degrees of freedom within the unit cell. Their tensor product acts on the internal degrees of freedom of the cell. The $\boldsymbol{k} = (k_x, k_y)$ is the 2D momentum. The $C_{x11}$ and $C_{y1}$ represent the intra cell jump intensity along the *x* and *y* directions and represents the inter cell coupling jump intensity. We set $C_{x12} = C_{y2} = C_0$ as the unit of energy for simplicity. Thus, the Hamiltonian in momentum space can be written as:

$$H(\boldsymbol{k}) = C_{x11} \tau_1 \sigma_0 + C_{y1} \tau_1 \sigma_1 - C_0 \cos k_x \tau_1 \sigma_3 + C_0 \sin k_x \tau_2 \sigma_4 + C_0 \cos k_y \tau_1 \sigma_1 - C_0 \sin k_y \tau_1 \sigma_2. \quad (3)$$

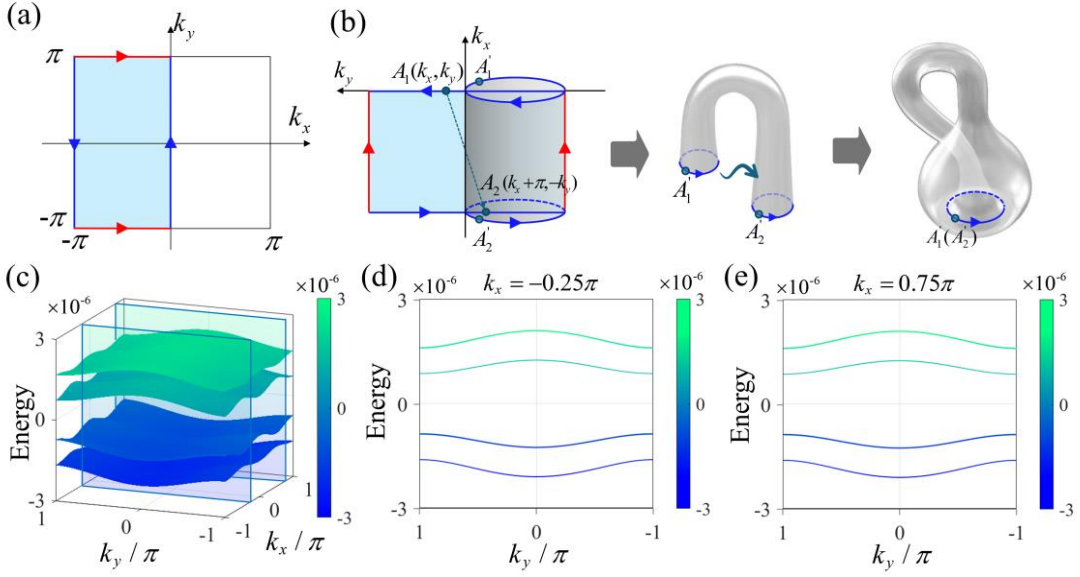

FIG. 2. Momentum space representation of the symmetry and Brillouin Klein bottle. (a) Fundamental domain of the BZ with boundaries of the same color identified along marked directions. (b) Cylinder with boundaries identified along opposite directions, forming the Brillouin Klein bottle. (c) Energy bands of the model in Fig. 1(b). (d) Cross-section of the band diagram in Fig. 2(c) for $k_y = -0.25\pi$. (e) Cross-section of the band diagram in Fig. 2(c) for $k_y = 0.75\pi$.

The system follows time-reversal (*T*) symmetry $TH(\boldsymbol{k})T^{-1} = H(-\boldsymbol{k})$ [21], where $T = \kappa$ (with $\kappa$ as the complex conjugate operator), as well as chiral symmetry $\Gamma H(\boldsymbol{k})\Gamma^{-1} = -H(\boldsymbol{k})$. The chiral symmetry operator is defined as $\Gamma = \tau_3\sigma_0$. Chiral symmetry serves as a protective mechanism, ensuring the quantization of energy associated with the quadrupole moment. For a system under the $\mathbb{Z}_2$ gauge field, we select one unit cell with the lattice constant $a_x$ along the *x*-direction. Fourier transformation shows that the translation symmetry $L_x$ is diagonalized as $L_x = e^{ik_x a_x}$. Therefore, $M_y e^{ik_x a_x} M_y = -e^{ik_x a_x} = e^{i(k_x + G_x/2)a_x}$ [23, 26, 27], where $G_x$ is the reciprocal lattice constant $G_x = 2\pi / a_x$. For simplicity, we typically set $a_x = 1$. Given the fractional reciprocal lattice constant in the exponent, it becomes apparent that $M_x$ acts as a glide reflection in momentum space:

$$M_y H(\boldsymbol{k}) M_y^{-1} = H(k_x + \pi, -k_y). \tag{4}$$

Then

$$H(k_x + \pi, -k_y) M_y |\psi(\boldsymbol{k})\rangle = E(\boldsymbol{k}) M_y |\psi(\boldsymbol{k})\rangle, \tag{5}$$

which indicates that $M_y|\psi(\mathbf{k})\rangle$ is an eigenstate of $H(k_x+\pi,-k_y)$. Hence, the eigenenergy at $(k_x,k_y)$ is equivalent to that at $(k_x+\pi,-k_y)$. Specifically, we further get the four eigenvalues of the Hamiltonian to plot its band structure with $k_x/\pi, k_y/\pi$ as the coordinate axes. It allows us to visually observe the glide symmetry in momentum space [23, 31], represented by $E(k_x,k_y)=E(k_x+\pi,-k_y)$. In Fig. 2(c), the blue rectangles outline two band slices of the bulk bands with a distance of π, depicted in Fig. 2(d) and (e) respectively. These two band structures exhibit opposing glide symmetry in momentum space.

Simultaneously, it should be noted that in the case of conventional space groups, nonsymmorphic symmetries such as glide reflections exclusively manifest themselves within real space lattices. In momentum space, they become fixed-point operations. There is always a momentum that remains invariant under these operations. However, the nonsymmorphic symmetry $M_y$ in momentum space discussed here operates without any fixed point. This type of symmetry, known as a free symmetry, can be further applied to reduce the BZ. As shown in Fig. 2(a), the fundamental domain shrinks to half of the BZ. The fundamental domain is chosen as $\tau_{1/2}=[-\pi,0)\times[-\pi,\pi)$ with periodic boundary conditions (PBC) for two (red lines) $k_y=-\pi,\pi$ edges and anti-PBC for two (blue lines) $k_x=-\pi,0$ edges. Due to $M_y H(\mathbf{k}) M_y^{-1}=H(k_x+\pi,-k_y)$, the orientations of the upper and lower boundaries of the cylinder are opposite. By bending and "sticking" them together, the Brillouin torus is converted into a Klein bottle.

The detailed procedure for the formation of a Klein bottle is illustrated in Fig. 2(b). The reduced BZ is curled into a cylinder. Due to $M_y H(\mathbf{k}) M_y^{-1}=H(k_x+\pi,-k_y)$, the orientations of the cylinder's upper and lower boundaries are opposite, with points $A_1$ and $A_2$ on the diagram mapping to $A_1'$ and $A_2'$ respectively. The cylinder is then bent to align the directions of the upper and lower boundaries, allowing them to be joined together, converting the Brillouin torus into a Klein bottle. Concurrently, the point $A_1$ undergoes a series of transformations and coincides with point $A_2$ on the Klein bottle. It is evident that the glider reflection can be observed from the relationship.

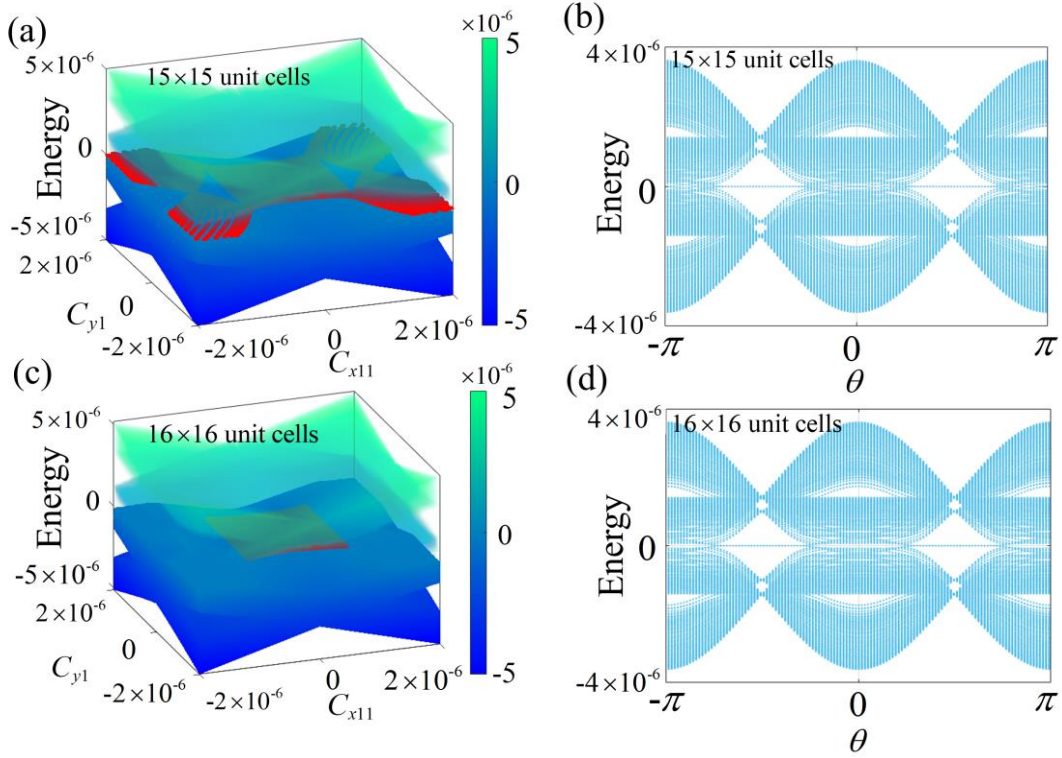

FIG. 3. Odd and even unit cell energy band diagrams. (a) The energy band diagram of the unit cells of 15×15. The red part represents the presence of zero energy states. (b) Scanning under the energy band diagram of 15×15 unit cells from $-\pi$ to $\pi$ in the case of $C_{x11} = 1.5 \times 10^{-6} \cos(\theta), C_{y1} = -1.5 \times 10^{-6} \cos(\theta)$. (c) The energy band diagram of the 16×16 unit cells. The red part represents the presence of zero energy states. (d) Scanning under the energy band diagram of 16 × 16 unit cells from $-\pi$ to $\pi$ in the case of $C_{x11} = 1.5 \times 10^{-6} \cos(\theta), C_{y1} = -1.5 \times 10^{-6} \cos(\theta)$.

Models of unit cells with odd and even numbers are constructed, and their band structures are shown in Fig. 3(a) and (c). The red regions indicate where the energy bands degenerate to zero, indicating the presence of zero-energy states. The specific properties of the red regions will be discussed later. To better illustrate the zero-energy states in the band structure, we make $C_{x11} = 1.5 \times 10^{-6} \cos(\theta), C_{y1} = -1.5 \times 10^{-6} \cos(\theta)$, and plot the band structure as a function of $\theta$ in Fig. 3(b) and (d). It can be seen that for unit cells with an odd number of quadrupoles, zero-energy states exist across the entire range from $-\pi$ to $\pi$. In contrast, for unit cells with an even number of quadrupoles, zero-energy states only exist in certain regions.

We investigate the Hamiltonian corresponding to $M \times M$ quadrupole units and its properties (corresponding to $N \times N$ nodes, where $N = 2M$). It helps to explain the differences in the energy bands of odd and even numbers of unit cells, as shown in Fig. 3. For the Klein bottle topoelectrical system, we can derive the Laplacian matrix $J$ of the circuit from the self-admittance and mutual admittance of each node, we further derive the circuit's Hamiltonian $H_{BBH}$, using $J = j\omega\left[\left(C_{01}+C_{02}-\dfrac{1}{\omega_0^2 L}\right)E + H_{BBH}\right]$. According to the relationship between eigenvalues and eigenvectors, we have the following equation $H_{BBH}|\psi\rangle = E|\psi\rangle$, then solve the energy bands [21]:

$$E = \pm\sqrt{\epsilon_y^2(k_y) + C_0^2 + C_{x11}^2 \pm 2C_{x11}\sqrt{\epsilon_y^2(k_y) + C_0^2 \cos^2 k_x}}, \tag{6}$$

where $\epsilon_y^2(k_y) = C_{y1}^2 + 2C_{y1}C_0 \cos k_y + C_0^2$. Translational invariance (by an even number of sites) suggests the following ansatz for an eigenstate of $H_{SSH}$

$$|\psi\rangle = \sum_{\beta=0}^{N_y-2}\sum_{\alpha=0}^{N_x-2} e^{-i(k_x\alpha + k_y\beta)}\left(A|4m+1\rangle + B|4m+2\rangle + C|4m+3\rangle + D|4m+4\rangle\right). \tag{7}$$

In Eq. (7), $\alpha$ and $\beta$ represent the $\alpha$-th and $\beta$-th nodes in $k_x$ and $k_y$ directions respectively.

By studying eigenvectors and constraint conditions as shown in Appendix A, we can conduct the subsequent analysis further. As $H_{BBH}$ is an $N^2 \times N^2$ Hamiltonian, it is expected that $N \times N$ solutions are required for a complete set. When $\mathbf{k}$ is real, the solutions for $|P|=0$ (the matrix $P$ is shown in Appendix A) are determined. In comparison with the complete set, the missing solution corresponds to an angular state. The solution of the equation $|P|=0$ is complicated. In order to observe the number and distribution of the solutions, we take a graphical method instead of directly solving the analytical solution, which is more convenient and intuitive. Specifically, we set $C_{x11}=0.47$ $uF$, $C_{y1}=0.33$ $uF$, and $C_0=1$ $uF$. Then, we observe the intersections of the real and imaginary parts of $|P|$ with the $xOy$ plane by plotting the real and imaginary parts of $|P|$ with $k_x$, $k_y$ as the $x$ and $y$ axes, respectively. The intersections point where both the real and imaginary parts of $|P|$ are zero correspond to the solutions $\mathbf{k}=(k_x,k_y)$ that satisfy the boundary constraints. The real and imaginary parts of $|P|$ is shown in Fig. 4(a) and (b) when $N=6$ ($M=3$, an odd number of quadrupole units). However, it is evident that for the real number $k$, there are merely 32 solutions depicted in Fig. 4(e), whereas there ought to be 36 solutions. All the 32 real solutions correspond to bulk states, since they exhibit oscillatory behavior as a function of the site index. The missing 4 solutions, corresponding to complex $k$, represent corner states. When $N=8$ ($M=4$, an even number of unit units), the real and imaginary parts of $|P|$ is shown in Fig. 4(c) and (d) and there should be 64 solutions. However, for the real $k$, only 62 solutions are corresponding to bulk states, and merely 2 missing complex solutions are corresponding to corner states, as can be observed in Fig. 4(f).

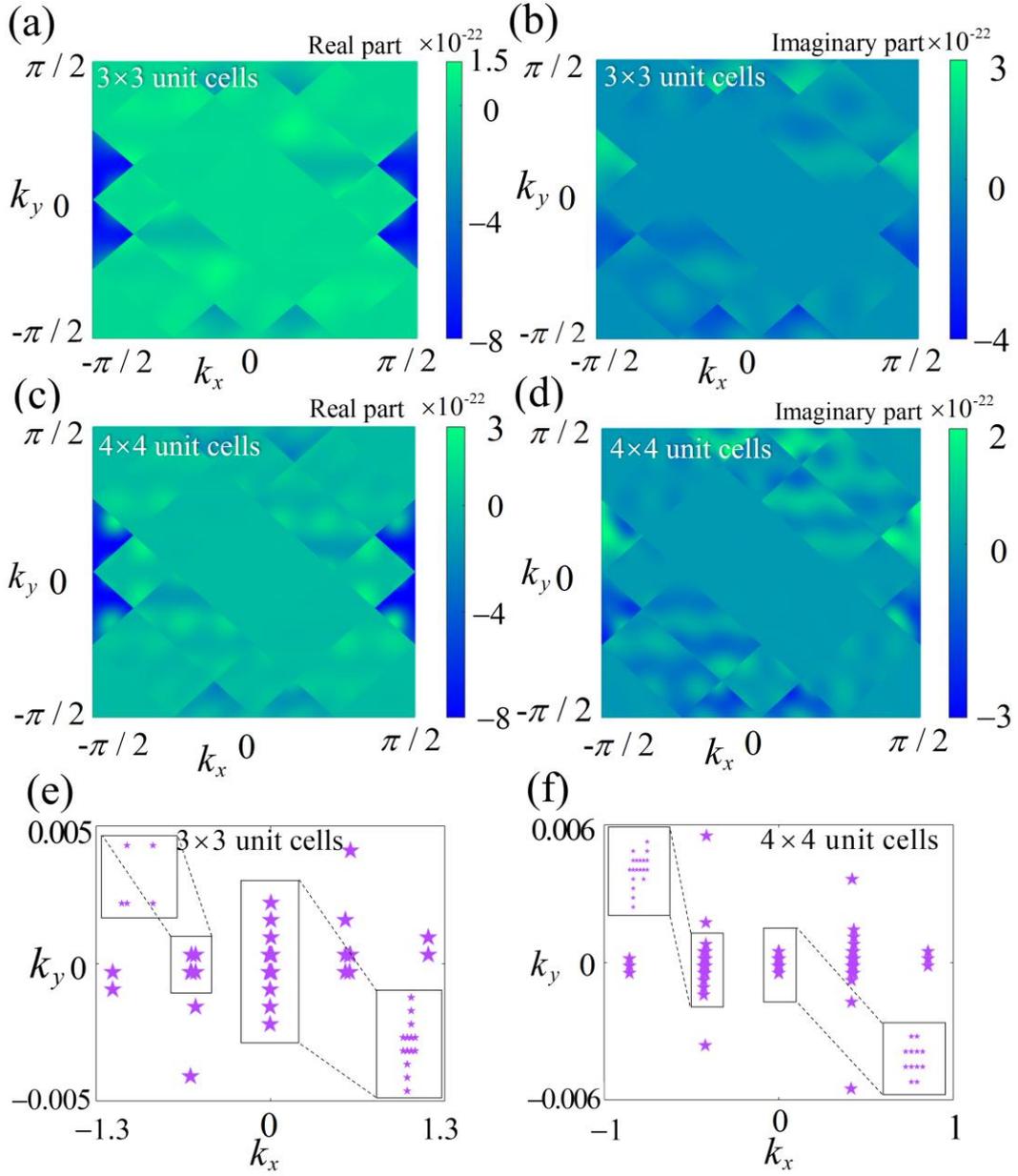

FIG. 4. (a) The real part of |P| at M=3. (b) The imaginary part of |P| at M=3. (c) The real part of |P| at M=4. (d) The imaginary part of |P| at M=4. (e) Distribution of real value solutions satisfying the |P| = 0 at M=3. (f) Distribution of real value solutions satisfying the |P| = 0 at M=4.

Furtherly, we plot the eigenvectors composed of odd and even numbers of unit cells in both the $k_x$ and $k_y$ directions. We select M=15 and M=16 (corresponding to N=30 and $N=32$). When the number of unit cells in the $k_x$ and $k_y$ directions is odd, the energy distributes across all four corners, indicating four corner states, as shown in Fig. 5. When the number of unit cells in the $k_x$ and $k_y$ directions is even, the energy is distributed in only two corners, indicating the presence of two corner states.

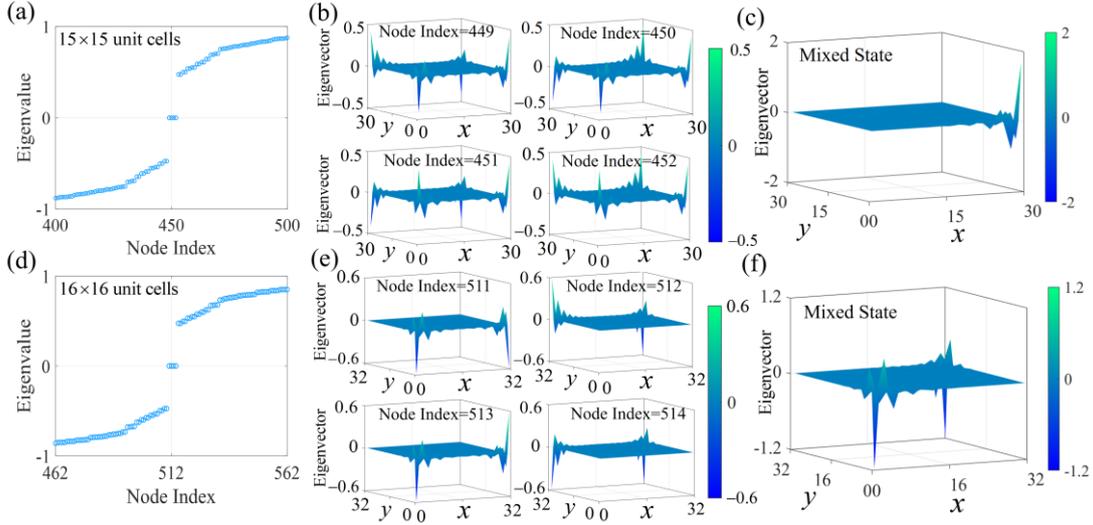

FIG. 5. Energy spectrum and mixed states in the Klein Bottle topoelectrical structure. (a) Energy spectrum for odd (15×15) unit cells under open bound-axies in both directions, exhibiting four zero-energy modes at the spectral center. (b) Mode localizations for the four zero-energy states, highlighting the localization characteristics at the corners of the odd (15×15) unit cells. (c) The mixed state is formed by superimposing four zero-energy modes under the odd (15×15) unit cells. (d-f) The same as (a-c) but for even (16×16) unit cells.

Then, we investigate the eigenstates of the Klein bottle topoelectrical system. The values of capacitance $C_{x11} = 470\ nF, C_{x22} = 470\ nF$, and $C_{y1} = 330\ nF$ play a significant role in localizing these eigenstates. The configuration of capacitance values has a zero-energy region in the energy band diagram of this topoelectrical system. The energy band diagram displays degenerate zero-energy modes, as illustrated in Fig. 5(a). The eigenvector shows their localized characteristics at the corners in Fig. 5(b). Because the topology is protected by spatial symmetry, corners create an effective edge for the boundary topological insulator and hence form the boundary of the boundary. In real-world circuits, the circuit parameters are impossible for them to be the same as the theoretical values due to the presence of errors in the components. It can result in their inability to locate at a certain eigenvalue. Because the zero-energy mode in Fig. 5(a) differs significantly from the other eigenvalues, we can consider that the actual circuit enters a mixed state of these four zero-energy modes. The mixed state formed by superimposing the four zero-energy modes in the odd (15×15) unit cells shows energy primarily localizing at a single corner, as depicted in Fig. 5(c).

In contrast, the even unit cell configuration displays distinct characteristics. The addition of an extra row and column in the lattice results in distinct symmetry conditions that influence the behavior of zero-energy modes. In the system consisting of 16×16 unit cells, the eigenvector reveals zero-energy modes spreading across two corners, as presented in Fig. 5(e). The corner states in Fig. 5(b) and (e) highlight the different topological localization mechanisms. The mixed state in the even unit cells, formed by combining zero-energy states, is depicted in Fig. 5(f). The eigenvector distributions for the four zero-energy states, pertaining to both odd and even numbered lattices, consistent closely with prior theoretical calculations. These corner state can potentially be realized using resonant circuits operating at specific frequencies.

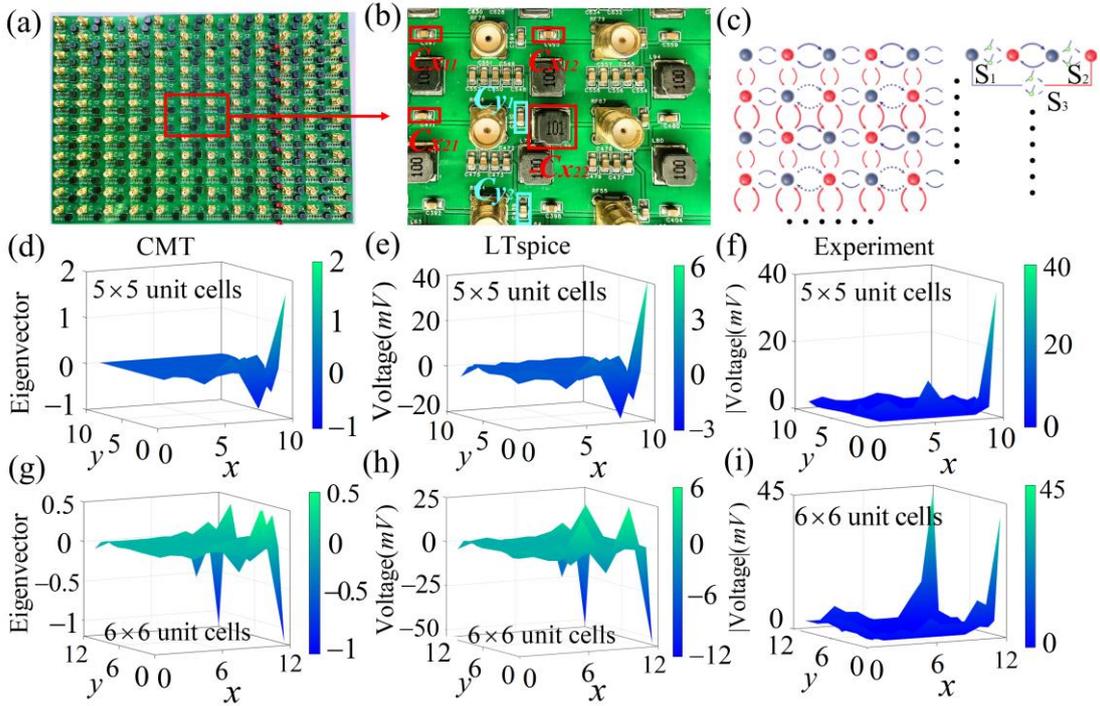

FIG. 6. Experimental circuit showcasing adjustable odd-even switching mechanism within a Klein bottle topological framework. (a) Comprehensive perspective of the experimental PCB. (b) Specific perspective of the experimental PCB. (c) Odd-even switching mechanism. (d) Distribution plots of characteristic vectors for odd unit cells by Coupled Mode Theory (CMT) calculations. (e) Circuit simulation outputs by LTspice for odd (5×5) unit cells. (f) Experiment outcomes from the circuit experimentation for odd unit cells. (g-i) The same as (d-f) but for even (6×6) unit cells.

In odd unit cells, the eigenvectors show zero energy modes spreading across one corner, while in even unit cells, the eigenvectors show zero energy modes spreading across two corners, highlighting distinct topological localization mechanisms. The LTspice simulations confirm the CMT-calculated corner state localization: one corner for odd unit cells and two corners for even unit cells. A 12×12 nodes PCB is designed to demonstrate these corner states, allowing transformation between 6×6 and 5×5 unit cells via jumper caps. The result validates the topological characteristics of the circuit and the phenomenon of corner states, as well as the previous theoretical calculations. The specific explanation of the experiment can be found in Appendix B.

In conclusion, we have successfully demonstrated the Klein bottle BBH topoelectrical model. Through the construction of a detailed topoelectrical model, we have explored the unique topological characteristics of the Klein bottle, particularly its glide reflection symmetry. Our theoretical calculation of odd and even unit cells has revealed distinct distributions of eigenvectors and localizations of zero-energy states. Experimental validation is conducted on a PCB using a switchable mechanism between odd and even configurations. Our work helps to bridge the gap between theoretical topological physics and applied electrical engineering, offering a unique platform for the exploration and realization of higher-order topological phases.

**Acknowledgements**

The authors thank for the support by NUPTSF (Grants No. NY220119, NY221055).

# APPENDIX A: Detailed derivation of the influence of odd and even number of quadrupoles on the number of corner states

The coupling relationship of the corresponding nodes in each quadrupole is consistent, which pertains to the case of the body state. However, this consistency does not apply to the nodes at the boundary for the Schrödinger equation $(H_{BBH} - E)|\psi\rangle = 0$. The system of Hamiltonian arises from the periodicity of the BZ in the $x$ and $y$ directions. Due to the difference of oscillation terms $e^{i2k_\zeta}$ ($\zeta = x, y$) between each unit the equations can always be simplified as follows

$$\begin{pmatrix} -E & (C_{x11} + C_{x12}e^{ik_x})^* & (C_{y1} + C_{y2}e^{ik_y})^* & 0 \\ C_{x11} + C_{x12}e^{ik_x} & -E & 0 & (C_{y1} + C_{y2}e^{ik_y})^* \\ C_{y1} + C_{y2}e^{ik_y} & 0 & -E & (C_{x21} - C_{x22}e^{ik_x})^* \\ 0 & C_{y1} + C_{y2}e^{ik_y} & C_{x21} - C_{x22}e^{ik_x} & -E \end{pmatrix} \begin{pmatrix} A \\ B \\ C \\ D \end{pmatrix} = 0.$$

(S1)

The Hamiltonian keeps constant under the phase shift $k_\zeta \to k_\zeta + \pi$ ($\zeta = x, y$), so we can set $k_\zeta$ in the range $(-\pi/2, \pi/2)$. It can be seen that there are four energy bands, according to the Eq. (6). Also, for a given energy solution, there are four solutions for $k$ in Eq. (6). Therefore, we need to use the same expression to take the linear combination of Eq. (7) and use A, B, C and D based on Eq. (S1) with $(k_x, k_y) \to (-k_x, k_y), (k_x, -k_y), (-k_x, -k_y)$ respectively, in order to obtain a complete system solution with $N \times N$ group.

$$|\psi\rangle = \sum_{j=0}^{N_y-2} \sum_{i=0}^{N_x-2} \{ \left(C_{++}A(k_x,k_y)e^{-i(k_xi+k_yj)} + C_{+-}A(-k_x,k_y)e^{-i(-k_xi+k_yj)} + C_{-+}A(k_x,-k_y)e^{-i(k_xi-k_yj)} + C_{--}A(-k_x,-k_y)e^{i(k_xi+k_yj)}\right)|4m+1\rangle$$
$$+ \left(C_{++}B(k_x,k_y)e^{-i(k_xi+k_yj)} + C_{+-}B(-k_x,k_y)e^{-i(-k_xi+k_yj)} + C_{-+}B(k_x,-k_y)e^{-i(k_xi-k_yj)} + C_{--}B(-k_x,-k_y)e^{i(k_xi+k_yj)}\right)|4m+2\rangle$$
$$+ \left(C_{++}C(k_x,k_y)e^{-i(k_xi+k_yj)} + C_{+-}C(-k_x,k_y)e^{-i(-k_xi+k_yj)} + C_{-+}C(k_x,-k_y)e^{-i(k_xi-k_yj)} + C_{--}C(-k_x,-k_y)e^{i(k_xi+k_yj)}\right)|4m+3\rangle$$
$$+ \left(C_{++}D(k_x,k_y)e^{-i(k_xi+k_yj)} + C_{+-}D(-k_x,k_y)e^{-i(-k_xi+k_yj)} + C_{-+}D(k_x,-k_y)e^{-i(k_xi-k_yj)} + C_{--}D(-k_x,-k_y)e^{i(k_xi+k_yj)}\right)|4m+4\rangle \}$$

(S2)

The $C_{++}, C_{-+}, C_{+-}$ and $C_{--}$ are the constants of the linear combination corresponding to the positive and negative solutions of $k_x$ and $k_y$.

On the other hand, the equations of the four corners of the system are taken to obtain the boundary constraints:

$$\begin{pmatrix} -E & C_{x11} & C_{y1} & 0 \\ C_{x11}e^{i(N_x-2)k_x} & -Ee^{-i(N_x-2)k_x} & 0 & C_{y1}e^{-i(N_x-2)k_x} \\ C_{y1}e^{i(N_y-2)k_y} & 0 & -Ee^{-i(N_y-2)k_y} & C_{x11}e^{-i(N_y-2)k_y} \\ 0 & C_{y1}e^{i[-(N_x-2)k_x+(N_y-2)k_y]} & C_{x11}e^{i[(N_x-2)k_x-(N_y-2)k_y]} & -Ee^{-i[(N_x-2)k_x+(N_y-2)k_y]} \end{pmatrix} \begin{pmatrix} A \\ B \\ C \\ D \end{pmatrix} = 0.$$

(S3)

Taking $\begin{pmatrix} A \\ B \\ C \\ D \end{pmatrix} = \begin{pmatrix} C_{++}A(k_x,k_y)+C_{-+}A(-k_x,k_y)+C_{+-}A(k_x,-k_y)+C_{--}A(-k_x,-k_y) \\ C_{++}B(k_x,k_y)+C_{-+}B(-k_x,k_y)+C_{+-}B(k_x,-k_y)+C_{--}B(-k_x,-k_y) \\ C_{++}C(k_x,k_y)+C_{-+}C(-k_x,k_y)+C_{+-}C(k_x,-k_y)+C_{--}C(-k_x,-k_y) \\ C_{++}D(k_x,k_y)+C_{-+}D(-k_x,k_y)+C_{+-}D(k_x,-k_y)+C_{--}D(-k_x,-k_y) \end{pmatrix}$

into Eq. (S3) and rearranging the equation, we get $P(C_{++},C_{-+},C_{+-},C_{--})^T = 0$, where

$$P = \begin{pmatrix} (-E \cdot A_{(++)}+C_{x11} \cdot B_{(++)}+C_{y1} \cdot C_{(++)})e^{i(2k_x+2k_y)} & (-E \cdot A_{(-+)}+C_{x11} \cdot B_{(-+)}+C_{y1} \cdot C_{(-+)})e^{i(-2k_x+2k_y)} & (-E \cdot A_{(+-)}+C_{x11} \cdot B_{(+-)}+C_{y1} \cdot C_{(+-)})e^{i(2k_x-2k_y)} & (-E \cdot A_{(--)}+C_{x11} \cdot B_{(--)}+C_{y1} \cdot C_{(--)})e^{-i(2k_x+2k_y)} \\ (C_{x11} \cdot A_{(++)}-E \cdot B_{(++)}+C_{y1} \cdot D_{(++)})e^{i(N_xk_x+2k_y)} & (C_{x11} \cdot A_{(-+)}-E \cdot B_{(-+)}+C_{y1} \cdot D_{(-+)})e^{i(-N_xk_x+2k_y)} & (C_{x11} \cdot A_{(+-)}-E \cdot B_{(+-)}+C_{y1} \cdot D_{(+-)})e^{i(N_xk_x-2k_y)} & (C_{x11} \cdot A_{(--)}-E \cdot B_{(--)}+C_{y1} \cdot D_{(--)})e^{-i(N_xk_x+2k_y)} \\ (C_{y1} \cdot A_{(++)}-E \cdot C_{(++)}+C_{x11} \cdot D_{(++)})e^{i(2k_x+N_yk_y)} & (C_{y1} \cdot A_{(-+)}-E \cdot C_{(-+)}+C_{x11} \cdot D_{(-+)})e^{i(-2k_x+N_yk_y)} & (C_{y1} \cdot A_{(+-)}-E \cdot C_{(+-)}+C_{x11} \cdot D_{(+-)})e^{i(2k_x-N_yk_y)} & (C_{y1} \cdot A_{(--)}-E \cdot C_{(--)}+C_{x11} \cdot D_{(--)})e^{-i(2k_x+N_yk_y)} \\ (C_{y1} \cdot B_{(++)}+C_{x11} \cdot C_{(++)}-E \cdot D_{(++)})e^{i(N_xk_x+N_yk_y)} & (C_{y1} \cdot B_{(-+)}+C_{x11} \cdot C_{(-+)}-E \cdot D_{(-+)})e^{i(-N_xk_x+N_yk_y)} & (C_{y1} \cdot B_{(+-)}+C_{x11} \cdot C_{(+-)}-E \cdot D_{(+-)})e^{i(N_xk_x-N_yk_y)} & (C_{y1} \cdot B_{(--)}+C_{x11} \cdot C_{(--)}-E \cdot D_{(--)})e^{-i(N_xk_x+N_yk_y)} \end{pmatrix}$$

(S4)

In the Eq. (S4), for the convenience of writing, we will abbreviate $A(k_x,k_y), A(-k_x,k_y), A(k_x,-k_y), A(-k_x,-k_y)$ as $A_{(++)}, A_{(-+)}, A_{(+-)}, A_{(--)}$, the same as $B$, $C$ and $D$. It should be noted here that $(C_{++},C_{-+},C_{+-},C_{--})^T = 0$ is an unphysical solution because wave functions $(C_{++},C_{-+},C_{+-},C_{--})^T$ become identically zero, i.e., electrons are absent in the system [39]. Therefore, $|P| = 0$ is required to meet the constraint conditions.

# APPENDIX B: Experimental details and testing instrument

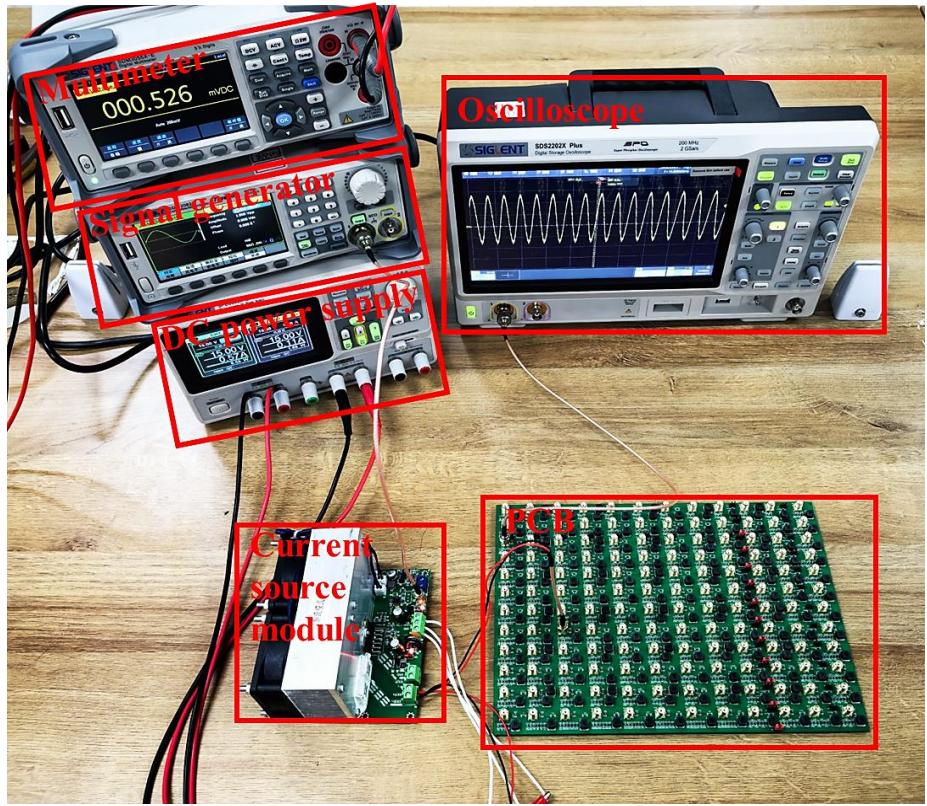

FIG. S1. Photos of PCB and testing instruments.

We employ JLC EDA software for PCB development, encompassing meticulous designs for PCB composition, stacking layout, inner layer configuration, and grounding design to mitigate signal crosstalk during experimental testing. To ensure precise measurements, we establish connections between the oscilloscope and signal generator using SMA interfaces and coaxial cables. Our laboratory is equipped with a comprehensive suite of SIGLENT instruments including the SDS2202X Plus Oscilloscope, SDG 2082X Function Waveform Generator, SPD3303X-E Programmable DC Power Supply, and SDM3055X-E Digital Multimeter. A voltage signal generated by the waveform generator is subsequently connected to an OPA549 voltage-controlled current source to serve as the excitation input for our experimental PCB. The OPA549 offers high voltage, high current, and excellent output swing capabilities, ensuring reliable performance for our experiments.

A 12×12 nodes PCB is constructed to elucidate the corner states of energy in an electrical circuit of the Klein bottle topology. Our circuit design ensures that its parameters are consistent with CMT. The complete image of the PCB is presented in Fig. 6(a). A magnified section of the PCB, displayed in Fig. 6(b), illustrates the coupling along the $x$ and $y$ axes. The PCB's built-in odd-even switching mechanism allows transformation between 6×6 and 5×5 unit cell circuits via jumper caps (Fig. 6(c)). Specifically, actuating switch $S_3$ to a closed state, while keeping $S_1$ and $S_2$ in an open state, morphs the circuit into 5×5 unit cell configurations. Conversely, engaging $S_1$ and $S_2$ to a closed state with $S_3$ open transfigures it into a 6×6 configuration. The flexibility of this circuit allows for effective modifications to test different topoelectrical configurations, offering an advantage over photonic and phononic crystal experiments.